\documentclass[11pt,a4paper,twoside,groupcitations]{article}
\usepackage[T1]{fontenc}
\usepackage[ansinew]{inputenc}
\usepackage[english]{babel}
\usepackage{amsfonts}
\usepackage{amsmath}
\usepackage{bm}
\usepackage{array}
\usepackage{amsthm}
\usepackage{amssymb}
\usepackage{graphicx}
\usepackage{subfigure}
\usepackage{braket}
\usepackage{eucal}
\usepackage{verbatim}
\usepackage[table]{xcolor}
\usepackage{caption}
\usepackage{cite}
\usepackage{textcomp}
\usepackage{multicol}
\usepackage{tikz}
\usetikzlibrary{positioning,arrows}
\usetikzlibrary{decorations.pathmorphing}
\usetikzlibrary{decorations.markings}
\usetikzlibrary{calc,decorations.markings}
\usetikzlibrary{arrows,shapes}
\usetikzlibrary{matrix,arrows}
\usepackage{pgfplots}
\usepackage{xparse}
\definecolor{jade}{HTML}{00A86B}
\newcommand{\be}{\begin{eqnarray}}
\newcommand{\ee}{\end{eqnarray}}

\renewcommand{\d}{\mbox{${\rm d}$}} 

\newcommand{\gn}{G_{\rm N}}

\def\beq{\begin{equation}}
\def\eeq{\end{equation}}
\def\@fmsl@sh#1#2#3{\m@th\ooalign{$\hfil#1\mkern#2/\hfil$\crcr$#1#3$}}
 \def\eq#1\en{\begin{equation}#1\end{equation}}
\def\s[#1,#2]{[#1\stackrel{\star}{,}#2]}
\def\sx[#1,#2]{[#1\stackrel{\star_{x}}{,}#2]}

\def\beq{\begin{equation}}
\def\eeq{\end{equation}}

\title{\bf Approximating compact objects in bootstrapped Newtonian gravity: use of the canonical potential}
\author{Roberto~Casadio$^{ab}$\thanks{E-mail: casadio@bo.infn.it},
$\ $
Iber\^e Kuntz$^{c}$\thanks{E-mail: kuntz@fisica.ufpr.br},
$\ $
and
Octavian Micu$^{d}$\thanks{E-mail: octavian.micu@spacescience.ro}
\\
\\
$^a${\em Dipartimento di Fisica e Astronomia, Universit\`a di Bologna}
\\
{\em via Irnerio~46, 40126 Bologna, Italy}
\\
\\
$^b${\em I.N.F.N., Sezione di Bologna, I.S.~FLAG}
\\
{\em viale B.~Pichat~6/2, 40127 Bologna, Italy}
\\
\\
$^c${\em Departamento de F\'isica, Universidade Federal do Paran\'a}
\\
{\em PO Box 19044, Curitiba -- PR, 81531-980, Brazil}
\\
\\
$^d${\em Institute of Space Science}
\\
{\em P.O. Box MG-23, RO-077125 Bucharest-Magurele, Romania}
}
\begin{document}
\maketitle
\begin{abstract}
We consider compact objects in a classical and non-relativistic generalisation of Newtonian gravity, dubbed bootstrapped
Newtonian theory, which includes higher-order derivative interaction terms of the kind generically present in the strong-field
regime of gravity.
By means of a field redefinition, the original bootstrapped Newtonian action is written in a canonical Newtonian form with
non-linear source terms. 
Exact analytic solutions remain unattainable, but we show that perturbative solutions of the canonical theory
can be efficiently used to derive approximate descriptions of compact objects.
In particular, using the canonical potential, we can more directly and generally show that the Arnowitt-Deser-Misner
mass differs from the (Newtonian) proper mass due to the non-linear couplings in the theory.
A few examples of sources with different density profiles are explicitly reanalysed in this framework.
\end{abstract}  
%
%
%
%
%
\newpage
\section{Introduction}
\label{sec:intro}
\setcounter{equation}{0}
Despite gravity being the oldest of the forces known to science, many gravitational phenomena and their quantum
foundations remain, to a large degree, fully open questions.
Roughly speaking, this can be mainly attributed to its weakness (as measured by the gravitational coupling
$\gn\sim 10^{-11}\,$m$^3\,$kg$^{-1}\,$s$^{-1}$).
When Newton's constant is combined with other ``small'' parameters (like $\hbar$ in a quantum regime),
gravitational effects seem to become utterly negligible at laboratory scales.
Testing aspects of strong gravity thus requires objects with huge masses (or large compactness) in order to produce
sizeable effects.
On the other hand, gravity is a non-linear phenomenon at its core and the theoretical study of compact objects
is a difficult endeavour due to the lack of general techniques to solve non-linear differential equations.
\par
Most of the known results are obtained perturbatively in the weak-field regime of general relativity,
far away from the compact source.
Perturbation theory indeed fails in the strong-field regime around compact objects, where an infinite tower
of couplings are generated in the non-relativistic approximation given by the expansion of the Einstein-Hilbert action.
In such a regime, all terms contribute equally and the infinite series cannot be truncated.
Inspired by this result, one could take a bottom-up approach and construct a modified
Newtonian theory by including, from the onset, terms of the functional forms which appear at the leading order in the
aforementioned expansion.
Rather than truncating the series obtained from general relativity, the resulting action is viewed as a new theory,
where finitely many terms are treated on the same foot.
From this perspective, the model functions as an alternative, rather than an extension, of Newtonian gravity,
in very much the same way that Stelle's higher-derivative gravity \cite{Stelle:1976gc} differs from general relativity.
The theory so obtained is called bootstrapped Newtonian gravity~\cite{Casadio:2018qeh}.
\par
One of the main purposes for devising the bootstrapped Newtonian gravity was to study static
(and spherically symmetric) compact sources~\cite{Casadio:2018qeh,Casadio:2019cux,Casadio:2020kbc,Casadio:2020ueb,Casadio:2019pli}.
A major difficulty however remains that the non-linearity of the field equation, and its interplay with the (Newtonian) conservation equation,
make it impossible to find analytical solutions.
It is therefore hard to derive general results to compare with the predictions of Newtonian physics or general
relativity.~\footnote{A major difference with respect to general relativity is the absence of a Buchdahl limit
for isotropic stars~\cite{Casadio:2019cux}.}
For this reason, in this work, we extend our previous investigations of compact sources in bootstrapped Newtonian
gravity by applying an idea introduced in Ref.~\cite{Casadio:2020mch}, whereby the kinetic term of the
bootstrapped Newtonian Lagrangian is put in canonical form by performing a field redefinition.
The role of the field redefinition is to replace derivative couplings, which are hard to deal with,
by standard (albeit non-linear) couplings. 
In particular, we shall employ the same Taylor expansion in powers of the radial coordinate $r$
from Refs.~\cite{Casadio:2018qeh,Casadio:2019cux} and show that the error one makes when truncating
the transformed solution (to second order in $r$) is negligible with respect to the truncated solution
before the field redefinition.
We stress that, because of the non-linear nature of the problem and of the required field redefinition,
this is a non-trivial result.
It is also practically relevant because working with the redefined canonical field makes it easier
to find approximate solutions for describing compact objects in the bootstrapped Newtonian theory.
\par
This paper is organised as follows:
in Section~\ref{bsgrav}, we review the bootstrapped Newtonian gravity and detail the aforementioned
field redefinition that brings the kinetic term to the canonical form;
in Section \ref{sec:genericJ}, we discuss approximate solutions that can be obtained for generic source terms
and some particular cases.
The main results we will obtain are that the bootstrapped Newtonian solutions are identical (at least) to second
order in $r$ and coincide with the Newtonian potential inside a homogeneous source, regardless of the 
actual density and pressure profiles;
terms with odd powers of $r$ must vanish and the difference with respect to the Newtonian potential 
appears at order $r^4$ (or higher);
the Arnowitt-Deser-Misner (ADM)-like mass~\cite{adm} $M$ and the (Newtonian) proper mass
$M_0$ are always different. 
Finally, we will draw some more conclusions in Section~\ref{sec:conc}. 
\section{Bootstrapped Newtonian gravity}
\setcounter{equation}{0}
\label{bsgrav}
We start by recalling that, in its most general form, the Lagrangian for the bootstrapped Newtonian potential $V=V(r)$
for static and spherically symmetric systems is given by~\cite{Casadio:2018qeh}~\footnote{We shall use units with $c=1$.}
\be
L[V]
&\!\!=\!\!&
L_{\rm N}[V]
-4\,\pi
\int_0^\infty
r^2\,\d r
\left[
q_V\,\mathcal{J}_V\,V
+
q_p\,\mathcal{J}_p\,V
+
q_\rho\, \mathcal{J}_\rho \left(\rho+q_p\,\mathcal{J}_p\right)
\right]
\nonumber
\\
&\!\!=\!\!&
-4\,\pi
\int_0^\infty
r^2\,\d r
\left[
\frac{\left(V'\right)^2}{8\,\pi\,\gn}
\left(1-4\,q_V\, V\right)
+\left(\rho+3\,q_p\,p\right)
V
\left(1-2\,q_\rho\, V\right)
\right]
\ ,
\label{LagrV}
\ee
where $L_{\rm N}$ is the Lagrangian for the Newtonian potential and $f'\equiv\d f/\d r$.
The motivation for each additional term was described extensively in previous
publications~\cite{Casadio:2018qeh,Casadio:2019cux,Casadio:2020kbc,Casadio:2020ueb,Casadio:2019pli},
so in the next paragraphs we will just briefly recall the meaning and role of each of them.
\par
The standard Newtonian Lagrangian, 
\be
L_{\rm N}[V]
=
-4\,\pi
\int_0^\infty
r^2 \,\d r
\left[
\frac{\left(V'\right)^2}{8\,\pi\,\gn}
+\rho\,V
\right]
\ ,
\label{LagrNewt}
\ee
yields the Poisson equation
\be
r^{-2}\left(r^2\,V'\right)'
\equiv
\triangle V
=
4\,\pi\,\gn\,\rho
\label{EOMn}
\ee
for the Newtonian potential $V=V_{\rm N}$ generated by the matter energy density $\rho=\rho(r)$.
As it was detailed in Refs.~\cite{Casadio:2018qeh,Casadio:2016zpl}, the gravitational self-coupling contribution
is then sourced by the gravitational energy $U_{\rm N}$ per unit volume, to wit
\be
\mathcal{J}_V
\simeq
\frac{\d U_{\rm N}}{\d \mathcal{V}} 
=
-\frac{\left[ V'(r) \right]^2}{2\,\pi\,\gn}
\ ,
\label{JV}
\ee
which couples to $V$ via the constant $q_V$ in Eq.~\eqref{LagrV}.
The static pressure $p=p(r)$ becomes very large for compact sources with a compactness~\cite{Casadio:2018qeh} 
\be
X
\equiv
\frac{\gn\, M}{R}
\gtrsim
1
\ ,
\label{defX}
\ee
where $M$ is the ADM-like mass that one would measure when studying
orbits~\cite{Casadio:2021gdf,DAddio:2021xsu} and $R$ is the radius of the source~\cite{Casadio:2019pli}. 
For this reason, a corresponding potential energy $U_p$ was added such that
\be
\mathcal{J}_p
\simeq
-\frac{\d U_p}{\d \mathcal{V}} 
=
3\,p
\ ,
\label{JP}
\ee
which couples to $V$ via the constant $q_p$ in Eq.~\eqref{LagrV}.
Since the above just adds to $\rho$, it can be easily included by simply shifting 
$\rho \to \rho+3\,q_p\,p$, where $q_p$ is a positive constant which formally allows us to implement the
non-relativistic limit as $q_p\to 0$.
Upon including these new source terms, and the analogous higher-order term $\mathcal{J}_\rho=-2\,V^2$,
which couples with the matter source, we obtain the total Lagrangian~\eqref{LagrV}.
\par
The Euler-Lagrange equation for $V$ is then given by
\be
\triangle V
=
4\,\pi\,\gn\left(\rho+3\,q_p\,p\right)
\frac{1-4\,q_\rho\,V}{1-4\,q_V\,V}
+
\frac{2\,q_V\left(V'\right)^2}
{1-4\,q_V\,V}
\ .
\label{EOMV}
\ee
We remark that the (dimensionless) coupling constants $q_V$, $q_p$ and $q_\rho$ track the effects of each additional
contribution and could be related to different specific theories of the interaction between gravity and matter
(for similar considerations, see, e.g.~Ref.~\cite{carloni}).
The Newtonian limit is clearly recovered for $q_V=q_p=q_\rho\to 0$.
\subsection{Field redefinition}
\label{subsec:redef}
The Lagrangian~\eqref{LagrNewt} can be generalised to non-static configurations $V=V(x^\mu)$ in
flat spacetime, thus yielding the kinetic term~\cite{Casadio:2020mch}
\be
K
=
-\left(1-4\,q_V\,V\right)
\frac{\partial_\mu V\,\partial^\mu V}{8\,\pi\,\gn}
\ ,
\ee
which is not in canonical form, and neither is $V$ of the canonical dimension for a scalar field.
In fact, we can change $K$ into the precise canonical form 
\be
K
=
-\frac{1}{2}\,\partial_\mu \psi\,\partial^\mu \psi
\ee
by means of the transformation~\cite{Casadio:2020mch}
\be
\psi
\equiv
\psi(V) 
=
\frac{1}{6\,\alpha}
\left[
1
-
\left(
1-4\,q_V V
\right)^{3/2}
\right]
\ ,
\label{psi_of_V}
\ee
where
\be
\alpha
=
q_V\,\sqrt{\gn}
\ .
\label{alpha}
\ee
The inverse relationship is given by
\beq
V
\equiv
V(\psi)
=
\frac{1}{4\,q_V}
\left[
1
-
\left(
1-6\,\alpha\,\psi
\right)^{2/3}
\right]
\ ,
\label{V_of_psi}
\eeq
and, after some algebra, the total Lagrangian~\eqref{LagrNewt} for static and isotropic configurations $\psi=\psi(r)$
reads
\be
L[\psi]
=
-
4\,\pi\int_0^\infty
r^2\,\d r
\left[
\frac{\left(\psi'\right)^2}{8\,\pi}
+
\left(J_{\rho}+3\,q_p\,J_{p}\right)
 \xi(\psi)
\right]
\ ,
\label{Lagrpsi}
\ee
in which the matter density was rescaled as
\be
J_{\rho}
=
\sqrt{\gn}\,\rho
\ ,
\label{Jrho}
\ee
like the pressure contribution  
\be
J_{p}
=
\sqrt{\gn}\,p
\ .
\label{Jp}
\ee
The interaction terms, which do not contain any derivatives of the new field $\psi$,
are all included in the non-linear coupling to the sources $J_\rho$ and $J_p$, that is
\be
\xi(\psi)
=
\frac{1}{4\,\alpha}
\left[1
-(1-6\,\alpha\,\psi)^{2/3}
\right]
\left\{1 
-
\frac{\beta}{2\,\alpha}\left[1
- \left(1-6\,\alpha\,\psi\right)^{2/3}\right]
\right\}
\ ,
\label{xi_psi}
\ee
where
\be
\beta
=
q_\rho\,\sqrt{\gn}
\ .
\label{beta}
\ee
\par
The most general form of the Euler-Lagrange equation for the canonically normalised $\psi=\psi(r)$ is
finally given by
\be
\triangle \psi
=
4\,\pi\,J\,
\frac{\alpha - \beta\left[1
- \left(1-6\,\alpha\,\psi\right)^{2/3}\right]}
{\alpha \left(1-6\,\alpha\,\psi\right)^{1/3}}
\ ,
\label{EOMpsi}
\ee
where $J\equiv J_{\rho}+3\,q_p\,J_{p}$ is the total effective density. 
In the following, we shall show that (approximate) solutions of Eq.~\eqref{EOMpsi} can indeed be
efficiently employed in order to determine (approximate) solutions of the original Eq.~\eqref{EOMV}.
\subsection{Vacuum solution}
\label{sec:vacuum}
In the vacuum outside a source of mass $M$ and radius $r=R$, we have $J=0$ and the above Eq.~\eqref{EOMpsi}
reduces to
\be
\triangle \psi
=
0
\ .
\ee
Of course, the exact solution for $r>R$ satisfying the proper asymptotic behaviour is the (canonically normalised)
Newtonian potential
\be
\psi_{\rm out}
=
-\frac{\sqrt{\gn}\,M}{r}
\ ,
\ee
which transforms back to~\cite{Casadio:2020mch} 
\be
V_{\rm out}
=
V(\psi_{\rm out})
\!\!\!\!&=&\!\!\!\!
\frac{1}{4\,q_V}
\left[
1
-
\left(
1+6\,q_V\,\frac{\gn\,M}{r}
\right)^{2/3}
\right]
\ .
\label{V_psi_N}
\ee
This is the exact solution of the bootstrapped Newtonian Eq.~\eqref{EOMV}
where $\rho=p=0$ with the expected asymptotic behaviour.
In particular, the Newtonian potential is recovered for $q_V\to 0$ and the first post-Newtonian
order of general relativity for $q_V= 1$~\cite{Casadio:2018qeh,Casadio:2021gdf,DAddio:2021xsu}.
\par
One important aspect that is not apparent from the above derivation is that the mass $M$ in Eq.~\eqref{V_psi_N}
is not equal to the proper mass $M_0$ of the source~\cite{Casadio:2018qeh,Casadio:2019pli}.
This follows precisely because of the non-linearity of Eq.~\eqref{EOMV} and the equivalent Eq.~\eqref{EOMpsi}.
We shall further investigate this aspect for the canonical field $\psi$ and various density profiles in the next
sections. 
\section{Quadratic approximation for the inner canonical potential}
\label{sec:genericJ}
\setcounter{equation}{0}

\par
To work with simpler equations, in the remainder of the paper we set $q_V=q_\rho$
(equivalent to $\alpha = \beta$), which simplifies the field equation~\eqref{EOMpsi} to
\be
\triangle \psi
=
4\,\pi\,J
\left(1-6\,\alpha\,\psi\right)^{1/3}
\ ,
\label{EOMpsi_gen}
\ee
where the effective density $J=J(r)$ vanishes outside the source of radius $r=R$. 
\par
Any solution of Eq.~\eqref{EOMV} for the bootstrapped Newtonian potential $V_{\rm in}=V(0\le r<R)$ needs
to match smoothly with the outer vacuum solution $V_{\rm out}$ in Eq.~\eqref{V_psi_N} across the boundary
$r=R$ of the source.
It is very easy to show that identical constraints must then hold for the field $\psi=\psi(V)$ satisfying
Eq.~\eqref{EOMpsi_gen}.
More precisely
\be
V_{\rm in}(R)
=
V_{\rm out}(R)
\equiv
V_R
\quad
\Leftrightarrow
\quad
\psi_{\rm in}(R)
=
\psi_{\rm out}(R)
\equiv
\psi_R \ ,
\label{bR}
\ee
and
\be
V'_{\rm in}(R)
=
V'_{\rm out}(R)
\equiv
V'_R
\quad
\Leftrightarrow
\quad
\psi'_{\rm in}(R)
=
\psi'_{\rm out}(R)
\equiv 
\psi'_R
\ ,
\label{dbR}
\ee
where we defined $\psi_{\rm in}=\psi(0\le r\le R)$ and $\psi_{\rm out}=\psi(R\le r)$.
Furthermore, we are looking for potentials generated by density profiles that are finite in the centre.
Therefore, the inner solution also needs to satisfy the regularity condition $V'_{\rm in}(0)=0$,
which in turn means that
\be
\psi_{\rm in}'(0)
=
0
\ .
\label{br0}
\ee
\par
Obtaining exact analytic solutions $V_{\rm in}$ for Eq.~\eqref{EOMV} or $\psi_{\rm in}$ for Eq.~\eqref{EOMpsi_gen}
is not feasible, even for an object with constant density and negligible pressure.
Hence, we will here focus on finding a good approximation by Taylor expanding around $r=0$.
Odd powers can be shown to vanish when imposing the constraint in Eq.~\eqref{br0}~\cite{Casadio:2018qeh}, 
so that the bootstrapped Newtonian potential up to second order reads
\be
V_{\rm in}
\simeq
V_0
+
V_2\,r^2
\ .
\label{V2}
\ee
This approximation was shown to work well for sources of small and intermediate compactness $X$
by comparing with numerical solutions in Refs.~\cite{Casadio:2018qeh, Casadio:2019cux}.
Therefore, we limit this case study to $X\lesssim 1$, which excludes objects hidden behind a horizon.
\par
The same Taylor expansion for the canonical potential reads
\be
\psi_{\rm in}
\simeq
\psi_0
+
\psi_2\,r^2
\ ,
\label{r2series}
\ee
and the mapping in Eq.~\eqref{V_of_psi} will then yield $\tilde V_{\rm in}=V(\psi_0+\psi_2\,r^2)$, which
we can compare with $V_{\rm in}$ in Eq.~\eqref{V2}.
Note that the two results will contain different powers of $r$ and their comparison is therefore not
straightforward.
However, one can estimate quantitatively how close the two approximations are to one another
by calculating their relative difference.
\subsection{General density and pressure profiles}
\label{subsec:genericJ}
Let us first consider a generic $J=J(r)$, with the only constraint that $J'(0)=0$.
Upon inserting the Taylor expansion~\eqref{r2series} in the equation of motion~\eqref{EOMpsi_gen} 
and employing the boundary condition~\eqref{dbR}, we find 
\be
\psi_0
=
\frac{1}{3\,\alpha}
\left[
1
-\frac{27\,\gn^{3/2}\,M^3}{R^9\,J^3_0}
\right]
\ee
where $J_0\equiv J(0)$, and
\be
\psi_2 
=
\frac{\sqrt{\gn}\,M}{2\,R^3}
\ .
\ee
From the matching condition in Eq.~\eqref{bR}, one obtains
\be
\psi_0
=
\frac{3\,\sqrt{\gn}\,M}{2\,R}
\ .
\ee
The approximate expression for the inner canonical field thus becomes
\be
\psi_{\rm in}
\simeq
-
\frac{\sqrt{\gn}\,M}{2\,R}\left(3 - \frac{r^2}{R^2}\right)
\ ,
\label{psi_in}
\ee
which is in fact the (canonically normalised) Newtonian solution for the inner potential inside a homogeneous source
satisfying 
\be
\triangle \psi
=
4\,\pi\,J_0
\ .
\label{EOMpsi_ap0_N}
\ee
This result shows that, as long as the quadratic approximations~\eqref{V2} and~\eqref{r2series} 
hold, the canonical $\psi_{\rm in}$ depends only on the mass $M$ and the size $R$ of the source.
From the inverse map~\eqref{V_of_psi}, we conclude that the bootstrapped Newtonian potential $V_{\rm in}$
is the same regardless of the density (and pressure) profile, up to terms of order $(r^4)^{2/3}$.
\par
We emphasise that, in the Newtonian case, the ADM mass $M$ equals the (Newtonian)
proper mass~\footnote{We notice that Eq.~\eqref{defM0} would give the ADM mass of a source in vacuum
in general relativity.
For this reason, we refer to $M_0$ as the Newtonian proper mass.}
\be
M_0
=
4\,\pi
\int_0^R
r^2\,\d r\,\rho(r)
\ .
\label{defM0}
\ee
In bootstrapped Newtonian gravity, $M$ and $M_0$ instead differ~\cite{Casadio:2018qeh,Casadio:2019pli},
since demanding that $M=M_0$ would over constrain the problem and yield no solution,
as will become apparent in the following subsections.
Calculating $M=M(M_0)$ is then very instructive, but it can only be done for given $J=J(r)$ and the 
corresponding expression can be very cumbersome.
\subsection{Homogeneous ball with negligible pressure}
\label{sec:zeropressure}
As a simple application, we directly consider a ball with homogeneous density 
\be
\rho
=
\rho_0\,\Theta(R-r)
=
\frac{3\, M_0}{4\,\pi\, R^3}\, 
\Theta(R-r)
\ ,
\label{HomDens}
\ee
where $\Theta$ is the Heaviside step function enforcing the density to vanish for $r>R$ and $M_0$
is the (Newtonian) proper mass defined in Eq.~\eqref{defM0}.
We also assume that the pressure be negligible, so that Eq.~\eqref{EOMpsi_gen} further simplifies to 
\be
\triangle \psi
=
4\,\pi\,J_{\rho}
\, (1-6\,\alpha\,\psi)^{1/3}
=
4\,\pi\,J_0
\, (1-6\,\alpha\,\psi)^{1/3}
\ ,
\label{EOMpsi_ap0}
\ee
where $J_0=\sqrt{\gn}\,\rho_0$, and we limit the investigation to values of the compactness $X\lesssim 1$,
as stated earlier. 
\subsubsection{Inner potential via field redefinition}
By plugging the Taylor expansion for the canonical $\psi_{\rm in}$ in Eq.~\eqref{r2series} into the equation
of motion~\eqref{EOMpsi_ap0}, we find
\be
\psi_2
\simeq
\frac{\sqrt{\gn}\,M_0}{2\,R^3}
\left(1-6\,\alpha\,\psi_0\right)^{1/3}
\ .
\label{PsiSer}
\ee
Eq.~\eqref{dbR} can next be used to determine
\be
\psi_0 
\simeq 
\frac{1}{6\,\alpha}
\left(1
- \frac{M^3}{M_0^3}
\right)
\ .
\label{psi0}
\ee
The matching condition~\eqref{bR} can be used to express the proper mass in terms of $M$, yielding
\be
\tilde M_0
=
\frac{M}{\left(1+9\,q_V\,\gn\,M/R\right)^{1/3}}
=
\frac{M}{\left(1+9\,q_V\,X\right)^{1/3}}
\ .
\label{M0psi}
\ee
Finally, we find the same approximate expression~\eqref{psi_in} corresponding to the Newtonian
solution of Eq.~\eqref{EOMpsi_ap0_N}. 
This shows that, at least in the quadratic approximation, the change in the coupling
\be
\xi(V)=1
\to
\xi(\psi)
=
(1-6\,\alpha\,\psi)^{1/3}
\ee
is equivalent to rescaling the Newtonian mass $M_0$ into the ADM mass $M$ according to Eq.~\eqref{M0psi}.
One could therefore solve the simpler Newtonian problem~\eqref{EOMpsi_ap0_N} and just write $M$ instead
of $M_0$ in the solution.
\par
\begin{figure}[t!]
\centering
\includegraphics[width=10cm,height=7cm]{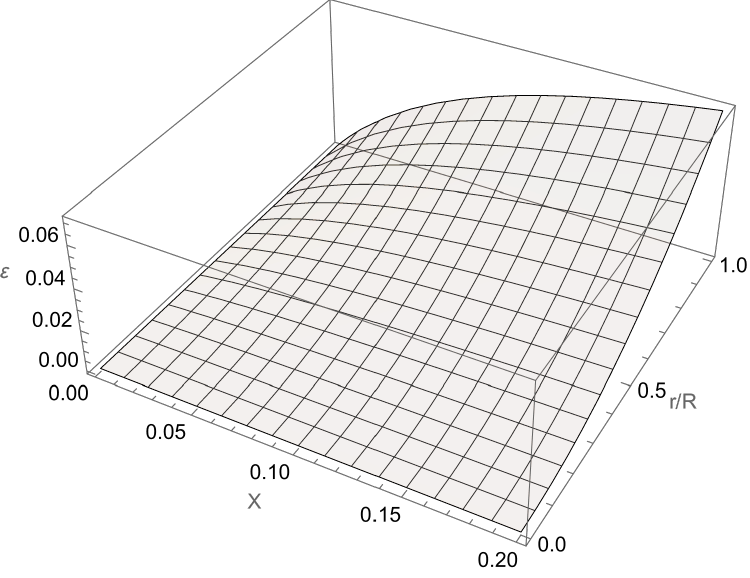}
\caption{Relative error from Eq.~\eqref{errPsi} for $q_V=1$.}
\label{errPsiplot}
\end{figure}

The bootstrapped Newtonian potential is obtained by substituting Eq.~\eqref{psi_in} in Eq.~\eqref{V_of_psi}
and reads
\be
\tilde V_{\rm in}
\!\!\!\!&\simeq&\!\!\!\!
\frac{1}{4\,q_V}
\left\{
1-
\left[1 + 3\,q_V\,X \left(3 - \frac{r^2}{R^2}\right)
\right]^{2/3}
\right\}
\nonumber
\\
\!\!\!\!&\simeq&\!\!\!\!
\frac{1}{4\,q_V}
\left\{
1-
\left(1 + 9\,q_V\,X \right)^{2/3}
\left[
1-\frac{2\,q_V\,r^2/R^2}{\left(1 + 9\,q_V\,X \right)}
\right]
\right\}
\ .
\label{tVin}
\ee
We can also estimate the relative error for this approximation by replacing the expression~\eqref{psi_in} into
Eq.~\eqref{EOMpsi_ap0}, from which we obtain
\be
\mathcal{E} \equiv \frac{q_V\,X}{1+9\,X}
\left(\frac{r}{R}\right)^2
\ll
1
\ ,
\label{errPsi}
\ee
which is displayed in Fig.~\ref{errPsiplot}.
Of course, this error vanishes everywhere inside the source in the Newtonian limit $q_V\to 0$ and is proportional
to the compactness $X$ otherwise.
\subsubsection{Inner bootstrapped Newtonian potential}
The approximate solution~\eqref{V2} for the homogeneous ball with vanishing pressure was found in Ref.~\cite{Casadio:2018qeh}
and is given by
\be
V_{\rm in}
\!\!\!\!&\simeq&\!\!\!\!
\frac{1}{4\,q_V}
\left[
1- \frac{1+2\,q_V\,X\left(4-r^2/R^2\right)}{\left(1+6\,q_V\,X\right)^{1/3}}
\right]
\nonumber
\\
\!\!\!\!&\simeq&\!\!\!\!
\frac{1}{4\,q_V}
\left\{
1-
\left(1 + 9\,q_V\,X \right)^{2/3}
\left[
1-\frac{2\,q_V\,r^2/R^2}{\left(1 + 9\,q_V\,X \right)}
\right]
\right\}
\ .
\label{sol}
\ee
The matching conditions across the surface also yield
\be
M_0
=
\frac{M}{(1+6\,q_V\,X)^{1/3}}\ ,
\label{M0M_V}
\ee
and one notices a different numerical factor multiplying $q_V$ in comparison with Eq.~\eqref{M0psi}.
\subsubsection{Comparing the approximations}
The main reason for this analysis is to understand if the field redefinition that brings the kinetic
term in a canonical form, thus simplifying the equation of motion,
leads to results that are in good agreement with those obtained without this transformation. 
The relative difference between the approximations in Eqs.~\eqref{sol} and \eqref{tVin} for small
compactness is given by
\be
\Delta
\equiv
\left|\frac{\tilde V_{\rm in}-V_{\rm in}}{V_{\rm in}}\right|
\simeq
\frac{q_V\,X}{6}
\left|
1-\frac{5\,r^2}{3\,R^2}
\right|
\ .
\label{Delta}
\ee
which is roughly of the same order as the error~\eqref{errPsi} shown in Fig.~\ref{errPsiplot}.
\begin{figure}[t]
\centering
\includegraphics[width=5cm,height=4cm]{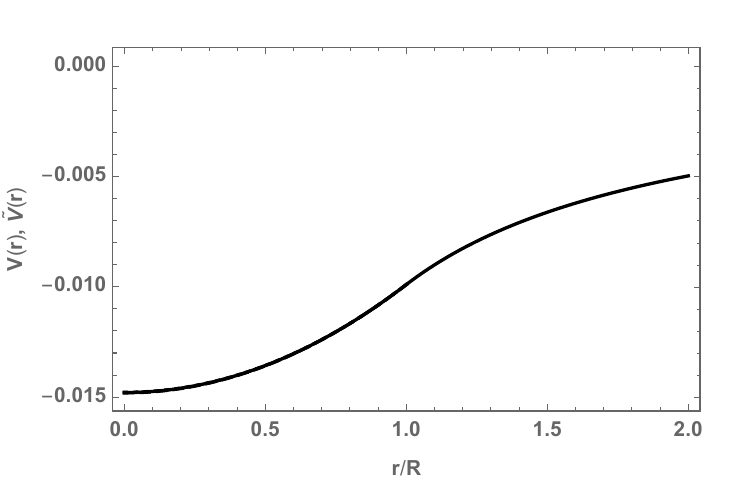}
$\ $
\includegraphics[width=5cm,height=4cm]{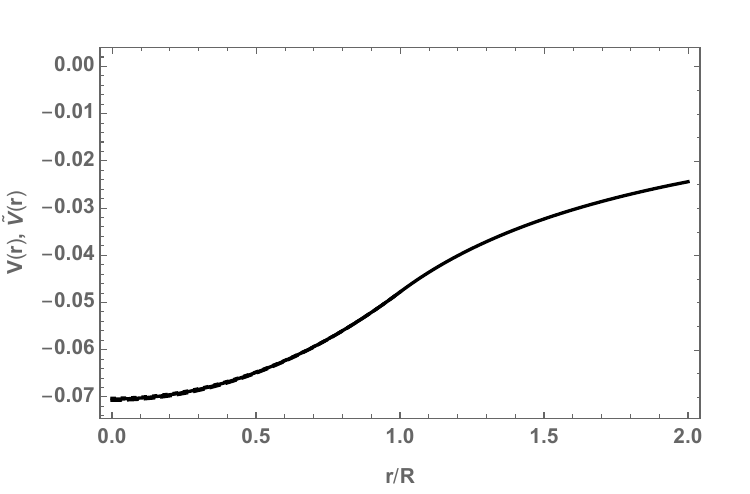}
$\ $
\includegraphics[width=5cm,height=4cm]{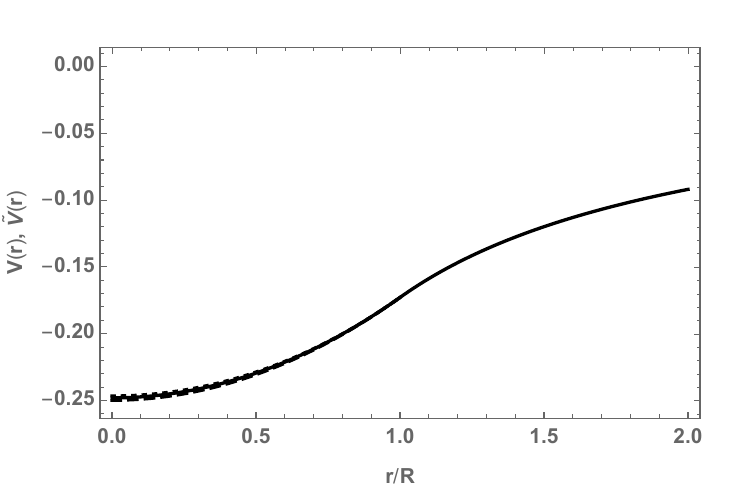}
\\
\includegraphics[width=5cm,height=4cm]{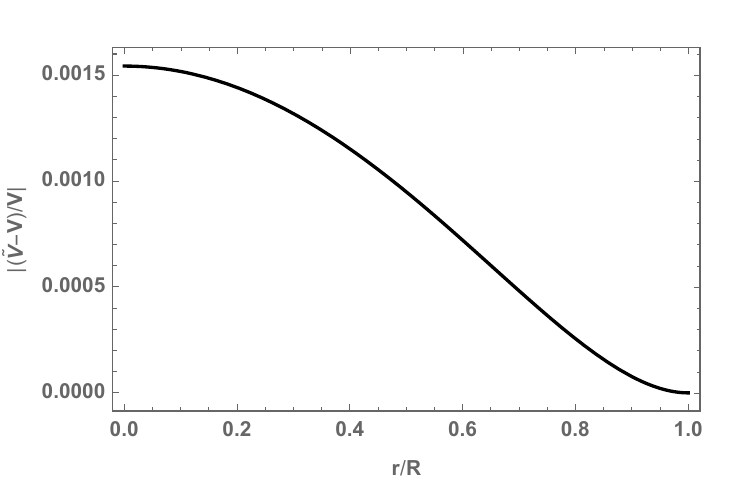}
$\ $
\includegraphics[width=5cm,height=4cm]{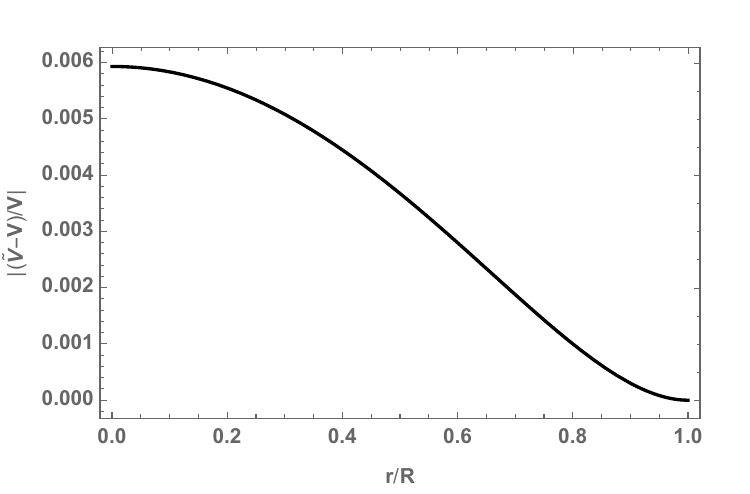}
$\ $
\includegraphics[width=5cm,height=4cm]{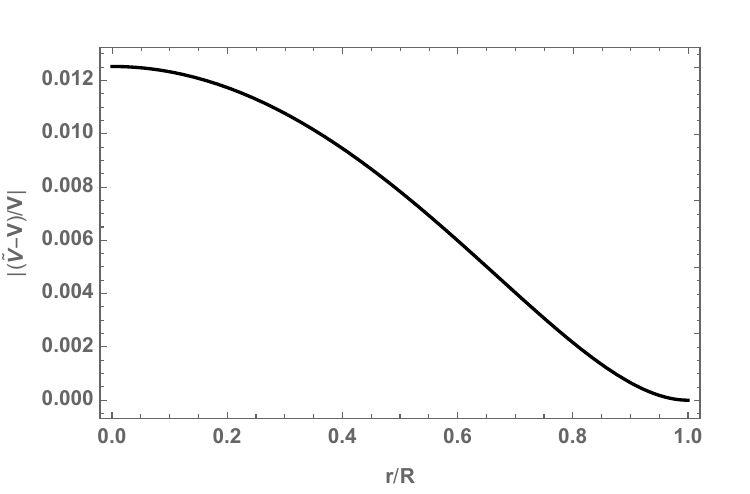}
\caption{\underline{Upper panels}:
inner potentials $\tilde V_{\rm in}$ (dotted line), $V_{\rm in}$ (dashed line), 
$\hat V_{\rm in}$ (solid line) and outer potential $\tilde V_{\rm out}$ (solid line) for $X =1/100$  (left panel),
$X = 1/20$ (center panel) and $X = 1/5$ (right panel).
\underline{Lower panels}:
absolute value of the relative difference $\Delta=|(\tilde V_{\rm in}-V_{\rm in})/V_{\rm in}|$ for the same compactness
as above. 
}
\label{V_comparison}
\end{figure}
\begin{figure}[t]
\centering
\includegraphics[width=5cm,height=4cm]{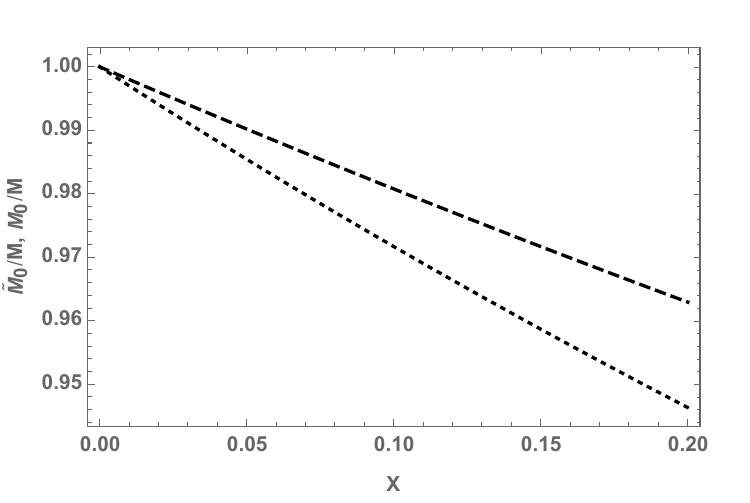}
$\ $
\includegraphics[width=5cm,height=4cm]{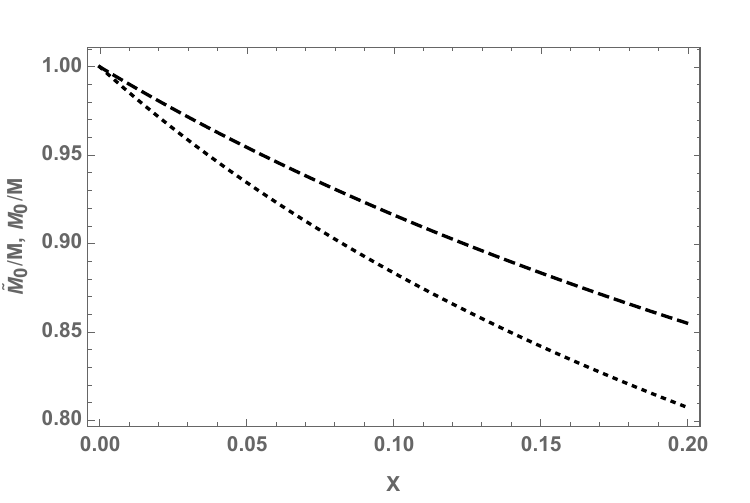}
$\ $
\includegraphics[width=5cm,height=4cm]{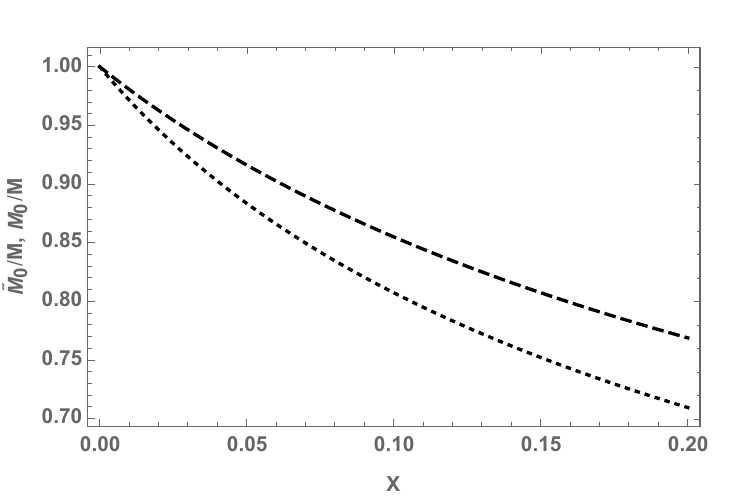}
\includegraphics[width=5cm,height=4cm]{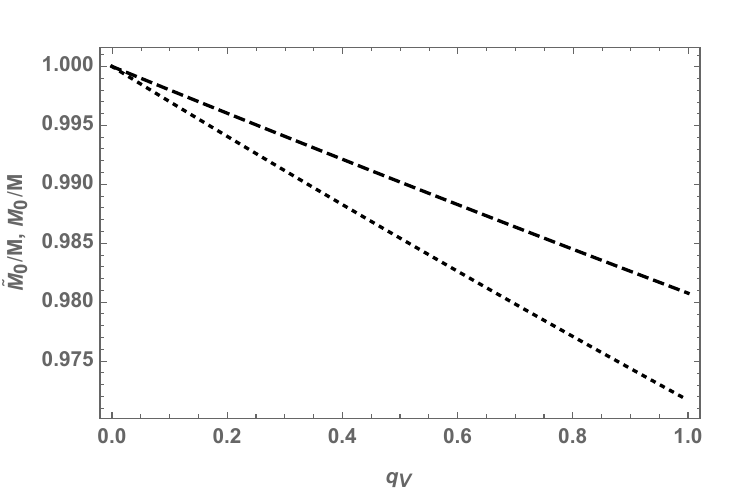}
$\ $
\includegraphics[width=5cm,height=4cm]{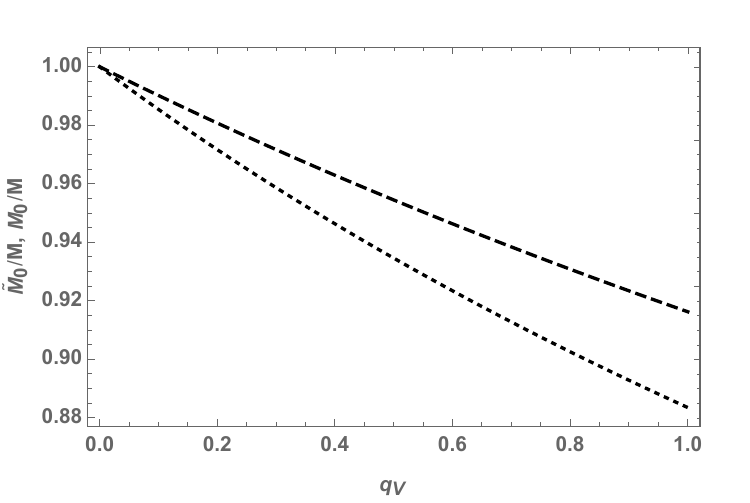}
$\ $
\includegraphics[width=5cm,height=4cm]{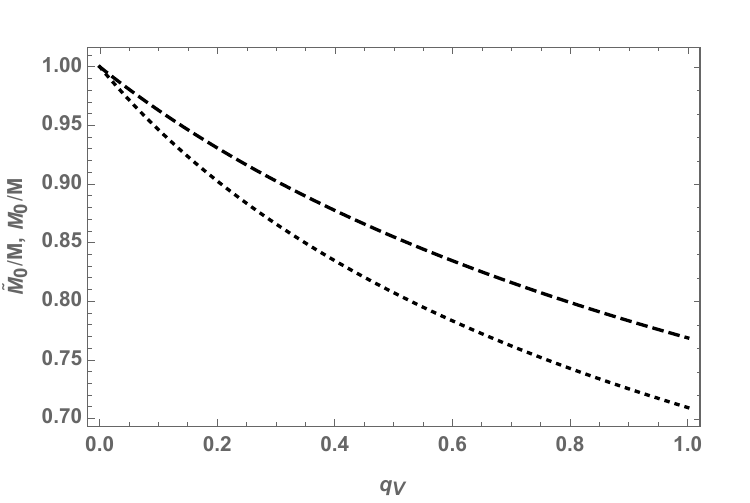}
\caption{\underline{Upper panels}:
ratios $\tilde M_0/M$ (dotted line) and $M_0/M$ (dashed line) as functions of $X$
for $q_V =0.1$  (left panel), $q_V =0.5$ (center panel) and $q_V =1$ (right panel). 
\underline{Bottom panels}: 
ratios $\tilde M_0/M$ (dotted line) and $M_0/M$ (dashed line) as functions of $q_V$
for $X =1/100$  (left panel), $X =1/20$ (center panel) and $X =1/5$ (right panel).
}
\label{M0M_plot}
\end{figure}
\par
Because of the non-linearity of the field equations, however, the above estimate remains of questionable
relevance.
In order to asses how reliable the analytical approximations are, we solve Eq.~\eqref{EOMV}
numerically with the same boundary conditions~\eqref{bR}-\eqref{br0} and
denote the numerical solution in the interior as $\hat V_{\rm in}$.
From the plot in Fig.~\ref{V_comparison}, we see that $\tilde V_{\rm in}$, $V_{\rm in}$ and $\hat V_{\rm in}$
follow each other very closely, both for small and intermediate values of the compactness. 
In the lower panels of the same figure one can also see plots of the relative difference
$\Delta$ for the same values of the compactness.
Therefore, even  though the approximate analytical expressions obtained for the potential
are different, both $\tilde V_{\rm in}$ and $V_{\rm in}$ appear to be in very good agreement
with the numerical results. 
\par
The main difference between the two approximate solutions $\tilde V_{\rm in}$ and $V_{\rm in}$
is how $\tilde M_0$ and $M_0$ depend on $M$.
Fig.~\ref{M0M_plot} shows the ratios $\tilde M_0/M$ and $M_0/M$ as functions of the compactness $X$
for the same values of $q_V$, respectively as functions of $q_V$ for the same values of $X$. 
\par
\begin{figure}[t]
\centering
\includegraphics[width=5cm,height=4cm]{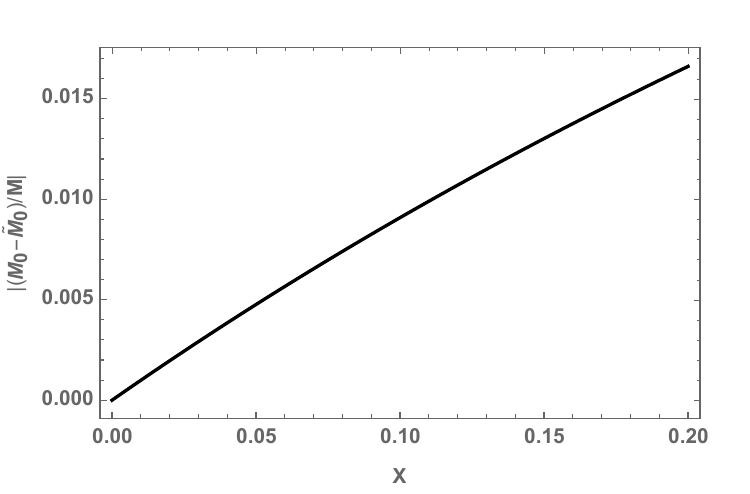}
$\ $
\includegraphics[width=5cm,height=4cm]{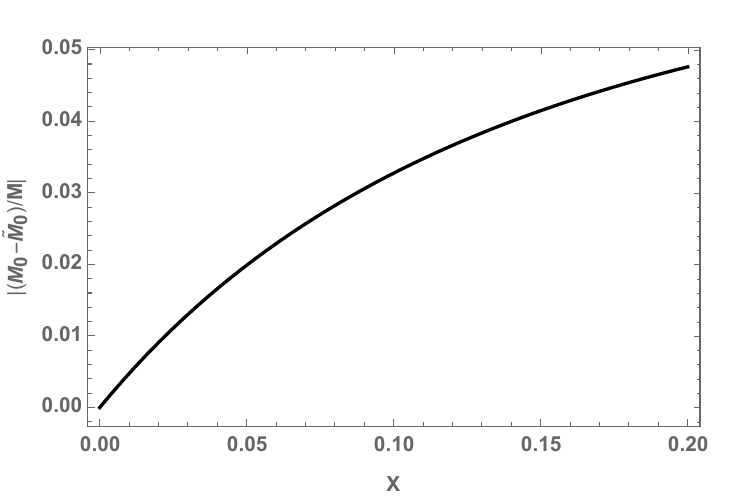}
$\ $
\includegraphics[width=5cm,height=4cm]{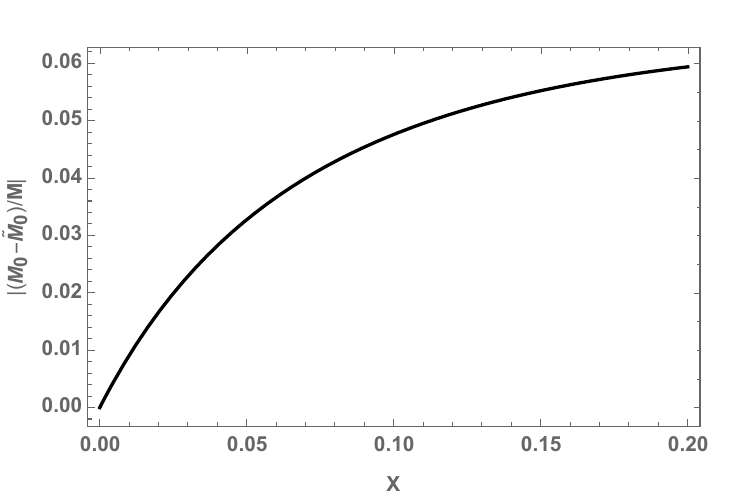}
\caption{Relative difference $|(\tilde M_0-M_0)/M|$ as a function of $X$ for $q_V =0.1$
(left panel), $q_V =0.5$ (center panel) and $q_V =1$ (right panel).}
\label{reldifM0_plot}
\end{figure}

The value of the coupling $q_V$ defines different regimes of the theory.
In the limit $q_V\to 0 $, one recovers Newtonian physics, as it is obvious from Eq.~\eqref{LagrV},
while the model is expected to approach General Relativity for $q_V\to 1$. 
This is why in Fig.~\ref{reldifM0_plot} we display $|(\tilde M_0-M_0)/M|$ for three different values of $q_V$.  
The relative difference increases with the compactness and with $q_V$.
Therefore, it is largest for $q_V=1$ for the largest compactness considered of $X=0.2$.
One notices that even in this case, the agreement remains fairly good. 
\subsection{Gaussian polytropic source}
\label{subsec:gaussian}
As a much less trivial example of relation between the ADM mass $M$ and the (Newtonian) proper mass $M_0$,
we consider a self-gravitating object described by the Gaussian density profile
\be
\rho
=
\rho_0\,e^{-\frac{r^2}{b^2\,R^2}} 
\,\Theta(R-r)
\ .
\label{rho_gaussian}
\ee
which was more extensively analysed in Ref.~\cite{Casadio:2020kbc}, using both numerical techniques and
analytical approximations.
This source becomes homogeneous for $b\gg 1$, while it is mostly concentrated around the centre for $b\ll 1$. 
We also assume a polytropic equation of state~\cite{horendt,Casadio:2020kbc} 
\be
p
=
\gamma\,\rho(r)
\left[\frac{\rho(r)}{\rho_0}\right]^{n-1} 
\equiv
\gamma\,\frac{\rho^n (r)}{\rho_0^{n-1}}
\ ,
\ee
where $\gamma$ and $n$ are the polytropic parameters
and the pressure clearly vanishes for $r>R$ due to Eq.~\eqref{rho_gaussian}.~\footnote{As was shown in Ref.~\cite{Casadio:2020kbc}, $p(R)>0$, but one can choose the polytropic parameters in such a way
that the pressure on the surface is negligibly small. Alternatively one could assume that a thin solid crust with a tension that balances the non-vanishing pressure covers the surface of the object.}
In this case, the (Newtonian) proper mass~\eqref{defM0} is given by
\be
M_0
=
\pi \,b^3\,R^3\,\rho_0
\left[\sqrt{\pi}\,{\rm Erf}\left(\frac{1}{b}\right)
- \frac{2}{b}\,e^{-1/b^2}
\right]
\ .
\label{pM0}
\ee
\par
In the quadratic approximation of Eq.~\eqref{r2series}, from the canonical field equation~\eqref{EOMpsi_gen} and
the matching condition~\eqref{dbR}, we find the new analytical approximation
\be
\psi_{\rm in}
\simeq
\frac{64\,\pi^3\,R^9\,\rho_0^3 \left(1+3\,\gamma \right)^3-27\,M^3}{384\,\pi^3\,\alpha\,R^9\,\rho_0^3 \left(1+3\,\gamma \right)^3}
+
\frac{\sqrt{\gn}\,M}{2\,R^3}\,r^2
\ ,
\label{2psi_rho}
\ee
where, from Eq.~\eqref{pM0}, the central density can be written in terms of $M_0$ as 
\be
\rho_0
=
\frac{M_0}{\pi \,b^2\,R^3
\left[\sqrt{\pi}\,b\,{\rm Erf}\left(1/b\right)
-2\,e^{-1/b^2}
\right] }
\ .
\label{M0_gauss}
\ee
The remaining matching condition~\eqref{bR} reads
\be
\frac{64\,\pi^3\,R^9\,\rho_0^3 \left(1+3\,\gamma \right)^3
-27\,M^3}
{384\,\pi^3\,\alpha\,R^9\,\rho_0^3 \left(1+3\,\gamma \right)^3}
= 
-\frac{3\,\sqrt{\gn}\,M}{2\,R}
\ ,
\label{b_gauss}
\ee
which shows that the solution~\eqref{2psi_rho} is indeed the same as the one in Eq.~\eqref{psi_in}.
Moreover, in this approximation, neither $\psi_{\rm in}$ nor the boundary condition~\eqref{b_gauss} depend on the
polytropic index $n$. 
\par
\begin{figure}[t]
\centering
\includegraphics[width=7cm,height=5.5cm]{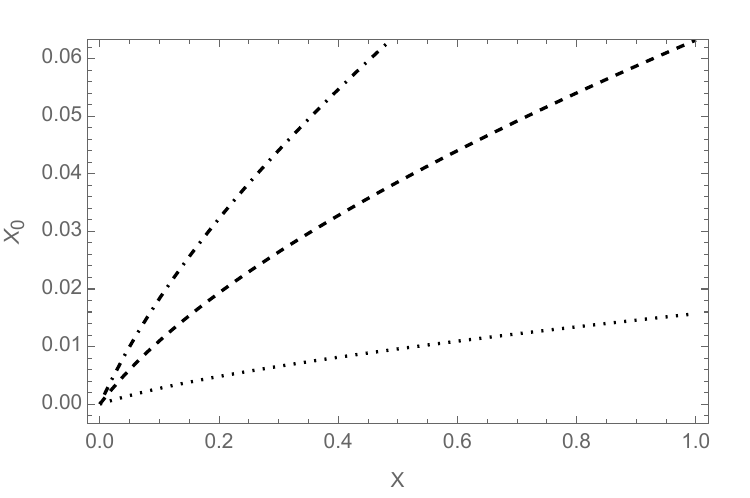}
$\ $
\includegraphics[width=7cm,height=5.5cm]{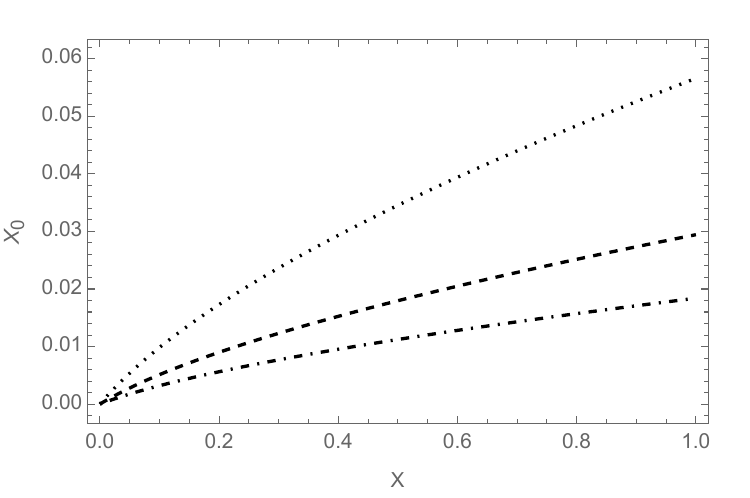}
\caption{Plots of $X_0$ as a function of $X$ for $q_V=1$.
\underline{Left panel}: 
$\gamma=0.5$, respectivelly $b=0.4$ (dotted), $b=0.7$ (dashed) and $b=1$ (dash-dotted). 
\underline{Right panel}: $b=0.5$, respectively $\gamma=0.1$ (dotted), $\gamma=0.5$ (dashed) and $\gamma=1$ (dash-dotted). 
}
\label{M_M0_plot}
\end{figure}

The expression~\eqref{b_gauss} can be written in terms of the compactness $X$ in Eq.~\eqref{defX} and
the analogue proper compactness $X_0=\gn\,M_0/R$ as 
\be
\frac{3}{2}\,X
=
\frac{27\,b^6\,X^3 \left[
\sqrt{\pi}\,b\,{\rm Erf}\left(1/b\right)
-2\,e^{-1/b^2}
\right]^3
-64\,X_0^3 \left(1+3\,\gamma \right)^3}
{384\,q_V\,X_0^3 \left(1+3\,\gamma \right)^3} 
\ ,
\label{b_gauss_1}
\ee
which shows the dependence of $X$ on $X_0$, equivalent to $M=M(M_0)$ for fixed $R$.~\footnote{This relation can be compared with the analogous Eq.~(4.10) of Ref.~\cite{Casadio:2020kbc}, which instead contains the index $n$.}
The plot of $X_0$ as a function of $X$ for several particular cases is shown in Fig.~\ref{M_M0_plot}. 
It is clear that $X_0<X$ in all displayed cases, which emphasises once more that $M$ and $M_0$
are generally different and setting them equal would only leave the trivial solution $M = M_0=0$.
Fig.~\ref{M_M0_plot} also shows that the ratio $X_0/X$ increases with the parameter
$b$ for constant $\gamma$, respectively decreases with $\gamma$ when $b$ is kept the same. 
\section{Conclusions}
\label{sec:conc}
\setcounter{equation}{0}
After having explored extensively the bootstrapped Newtonian gravity model in a series of previous papers,
we have now tested an alternative approach to finding solutions for various cases within the same model.
In its standard form, besides the Laplacian, the Euler-Lagrange equation for the bootstrapped potential
contains extra derivatives of the potential.
This makes it very cumbersome (when at all possible) to obtain solutions, in addition to hindering the true
degrees of freedom of the theory.
After performing a field redefinition, one can write the theory in canonical form, which is easier to solve
and has a more transparent interpretation.
\par
In order to test the effectiveness of the new formulation in terms of the canonical potential,
we solved the canonical equation of motion~\eqref{EOMpsi} for a general density profile in a quadratic approximation
around the centre, and compared with the results obtained by solving the non-canonical equation. 
We emphasise that the field redefinition~\eqref{V_of_psi} allowed us to prove some general results:
in the approximation~\eqref{V2}, at least up to terms of order $(r^4)^{2/3}$,~\footnote{The power $4$ is from
the Taylor expansion and the power $2/3$ from the field transformation~\eqref{V_of_psi}.}
the interior bootstrapped Newtonian potential does not depend on the density or pressure profile of the source,
but only on its mass and radius.
The density profile will, however, determine the relationship between the (Newtonian) proper mass $M_0$
and the ADM mass $M$ of the source.
The other striking and seemingly general property observed in all cases is that these two masses
are different, $M_0 \neq M$, regardless of the values for any other parameters.
This is a fundamental difference with respect to Newtonian gravity and is indeed expected
due to the non-linear nature of the theory. 
Nonetheless, we remark that the Newtonian $M_0$ in our approach is different from the proper mass in general
relativity, so that the discrepancy we find between $M$ and $M_0$ could lead to important consequences
in cosmology and astrophysics.
Some of these phenomenological applications will be considered throughly in a separate paper~\cite{pheno}.
\par
We should also stress that, in principle, field redefinitions cannot change the physical content of a (classical)
theory.
Nevertheless, when one considers non-linear redefinitions, such as Eq.~\eqref{V_of_psi} employed in this paper,
approximate solutions of the canonical equations are not in one-to-one correspondence with the
approximate solutions of the non-canonical ones.
In fact, truncating the solution of either side of the equivalence would require infinitely many terms
from the other side.
One should thus expect large errors when comparing truncated solutions of each side of the correspondence.
We, however, showed that such an error is quite small for the regimes we are interested in,
thus opening up the possibility of solving the equations of motion in a much simpler way. 
\section*{Acknowledgments}
R.C.~is partially supported by the INFN grant FLAG and his work has also been carried out in
the framework of activities of the National Group of Mathematical Physics (GNFM, INdAM).
O.M.~is supported by the Romanian Ministry of Research, Innovation, and Digitisation, grant no.~16N/2019
within the National Nucleus Program.
%
%
%
%

%
\end{document}